\documentclass[12pt,aps]{revtex4}

\usepackage{graphics}
\usepackage{epsfig}

\newcommand{\beq}{\begin{equation}}  
\newcommand{\eeq}{\end{equation}}
\newcommand{\bea}{\begin{eqnarray}}
\newcommand{\eea}{\end{eqnarray}}

\begin{document}

\title{Radiative seesaw: a case for split supersymmetry}

\author{Borut Bajc$^{(1,2)}$ and Goran Senjanovi\' c$^{(2)}$}
\affiliation{$^{(1)}$ {\it J. Stefan Institute, 1001 Ljubljana, Slovenia}
\footnote{Permanent address}}
\affiliation{$^{(2)}${\it International Center for Theoretical Physics, 
Trieste, Italy}}

\begin{abstract} 
We revive Witten's mechanism for the radiative seesaw induced neutrino
masses in SO(10) grand unified theory. We propose its extension to charged
fermion masses as a possible cure for wrong tree level mass relations. 
We offer two simple realizations that can produce a realistic fermionic
spectrum. The first one requires two 10 dimensional Higgses in the Yukawa
sector and utilizes radiative effects for charged fermion masses. The 
second one trades one $10$ for a $120$ dimensional Higgs and leads to the 
SO(10) theory with less parameters in the Yukawa sector. The mechanism 
works only if supersymmetry is broken at the GUT scale while gauginos 
and higgsinos remain at TeV. This provides a strong rationale for the 
so called split supersymmetry.
\end{abstract}


\maketitle


\section{Introduction}

The simplest and the most popular approach for generating small neutrino 
masses is based on the seesaw mechanism \cite{Minkowski:1977sc}. This 
scenario must be implemented in a well defined theory in order to 
be predictive and testable. The SO(10) grand unified theory provides 
a natural framework since it contains automatically righthanded neutrinos 
and due to unification constraints restricts the seesaw scale. With the 
advent of neutrino masses it can be thus argued that SO(10) is actually 
the minimal realistic grand unified theory. In this context SU(5), which 
was tailor made for massless neutrinos, becomes cumbersome and ridden 
by too many parameters. 

Among a number of different ways of realizing the seesaw mechanism 
in SO(10), the one of Witten \cite{Witten:1979nr} stands out for its 
simplicity and beauty. It is based on two-loop radiatively induced 
and calculable righthanded neutrino masses if the $B-L$ symmetry is 
broken by a $16$ dimensional Higgs multiplet. We call it the 
radiative seesaw mechanism. Obviously it must fail 
in any low energy supersymmetric theory due to the nonrenormalization 
theorem of the superpotential. Since in the last two decades most of the 
effort went into supersymmetric grandunification, this appealing approach 
unfortunately fell from grace. Still, this mechanism is too 
appealing to be given up. In this letter we revive this approach 
and, equally important, we extend it to the charged fermion masses. 
In the process we suggest two simple minimal realizations that could 
lead to realistic theories. 

Regarding low-energy supersymmetry, its main motivation 
is the control of the gauge hierarchy in perturbation theory. 
If one is to accept the fine-tuning of the Higgs mass the way one does 
for the cosmological constant, the scale of supersymmetry breaking becomes 
a dynamical issue to be determined by the unification constraints. In 
principle in SO(10) gauge coupling unification needs no supersymmetry at 
all provided there is an intermediate scale \cite{Chang:1984qr}. Actually, 
even without supersymmetry there is a very appealing mechanism of 
understanding a gauge hierarchy based on the attractor vacua 
\cite{Dvali:2003br}. The need for low-energy supersymmetry disappears 
also in the landscape picture where one stops worrying about 
the smallness of the weak scale. This fits nicely with the anthropic 
arguments in the favour of the small cosmological constant 
\cite{Weinberg:1987dv}. 

On the other hand, if one abandons the need for the perturbative stability
of the Higgs mass, grandunification does not tell us what the effective
theory at TeV energies relevant for LHC is. This is the burning question
and any guidance is badly needed. In the minimal SU(5) theory the options
are limited: unification constraints require either low-energy
supersymmetry or split supersymmetry with light gauginos and higgsinos and
heavy sfermions \cite{Arkani-Hamed:2004fb,Giudice:2004tc}.

The trouble as we said is that SU(5) is not 
a good theory of neutrino masses and furthermore, it cannot decide 
between the supersymmetric and the split supersymmetric options above. 

The motivation behind this letter is twofold. We wish to construct a 
simple and predictive realistic theory based on the radiative generation 
of neutrino masses and, at the same time, we would like to determine the 
effective low-energy theory. Obviously, as we said, the low-energy theory 
cannot be supersymmetric, since the righthanded neutrino mass would then 
be suppressed by the small scale of supersymmetry breaking. We will show 
though, that the phenomenological and unification constraints lead 
automatically to split supersymmetry. This provides a strong motivation 
for a large scale of supersymmetry breaking. The LSP dark matter is then 
a welcome consequence rather than an input as in the original work.

The original model of Witten utilized a single $10$ dimensional Higgs 
and ended up predicting neither quark nor leptonic mixings and the usual 
bad mass relations $m_s=m_\mu$ and $m_d=m_e$ at the GUT scale. Even worse, 
neutrino masses scale as up quark masses. The failure does not lie in the 
radiative mechanism of the righthanded neutrino mass, but rather in the 
oversimplistic Yukawa Higgs sector. In order to get a correct mass spectrum 
of charged fermions one must complicate the Yukawa sector. One possibility 
is adding a $\overline{126}_H$ dimensional Higgs representation, which 
then works 
successfully at the tree level. This has been worked out in detail in 
the context of supersymmetric SO(10) \cite{Babu:1992ia}, but can equally 
well be implemented in the nonsupersymmetric version \cite{toappear}. 
In the radiative mechanism case one should instead add another $10_H$ 
or $120_H$. In this work we discuss both versions and show how they 
promise to offer realistic theories of fermion masses and mixings. 
It may appear impossible to have a realistic theory with two $10$'s 
due to the fact that the above bad relations apparently do not depend on 
the number of these multiplets. This is not true though once we go 
beyond the tree level. We find that the Witten's radiative approach 
is readily generalized to light fermions. 

The two $10_H$'s version is appealing since the charged fermion masses 
are corrected radiatively, whereas the version with $120$ is attractive 
due to the smaller number of parameters. 

\section{The model}


The natural theory to start with is the one with $16_H$ (and, 
normally in supersymmetry one takes also $\overline{16_H}$) 
and $45_H$ Higgses. This however is not enough, since it can be shown 
that it leaves SU(5) unbroken \cite{Li:1973mq}. One can simply add a 
$54_H$ or use $210_H$, which works by itself. The choice is not so 
important for $\nu_R$; what is crucial is to use the $16_H$. 
It may be relevant though for radiatively induced corrections to light 
fermion masses (see model B below). Either choice leaves the rank 
unbroken, i.e. at least a $B-L$ symmetry remains intact (usually also 
SU(2)$_R$ remains a good symmetry). The next stage of symmetry breaking 
is achieved by $\langle 16_H\rangle=M_R$. Whether or not $M_R$ lies at 
$M_{GUT}$ is determined by the unification and phenomenological 
constraints. In this theory one ends up with a single step breaking, i.e. 
$M_R=M_{GUT}\approx 10^{16}$ GeV, due to the neutrino mass considerations. 
This is discussed below. 

On top of that we need a ``light'' Higgs responsible for the electroweak 
scale. The simplest and the most common choice is a $10_H$ dimensional 
multiplet with the Yukawa interaction schematically

\beq
{\cal L}_{Y}=16_FY_{10}10_H16_F\;.
\eeq

\begin{figure}[h]
\includegraphics[width=0.70\linewidth]{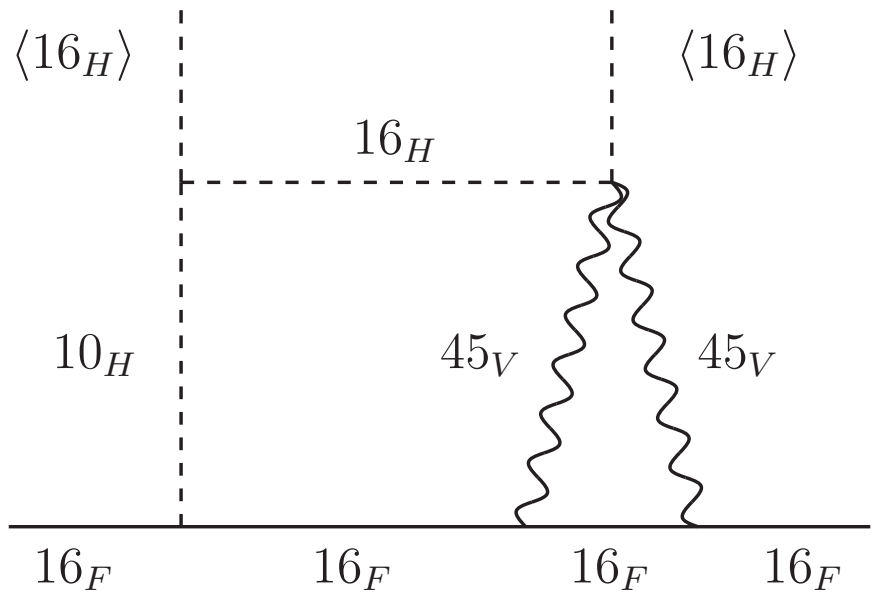}
\caption{\label{fig1} A contribution to the radiatively 
generated fermion mass.}
\end{figure}

As is well known, righthanded neutrino masses, being SU(5) singlets, 
can only arise from a five index antisymmetric $\overline{126}$ 
representation, missing in this approach. In the language of the 
SU(2)$_L\times$SU(2)$_R\times$SU(4)$_C$ Pati-Salam symmetry 
(hereafter denoted as PS) one needs a nonzero vev in the 
$(1,3,10)$ direction. Thus it must be generated radiatively 
and it can only appear at the two loop level shown in Fig. \ref{fig1}. 

One obtains \cite{Witten:1979nr}

\beq
\label{mnur}
M_{\nu_R}\approx\left({\alpha\over\pi}\right)^2Y_{10}
{M_R^2\over M_{GUT}}\;.
\eeq

Notice that we write $M_R^2/M_{GUT}$ instead of $M_{GUT}$ in 
\cite{Witten:1979nr} in order to be as general as possible. 
Of course this was a nonsupersymmetric theory. Today we 
know that this must fail as mentioned in the introduction. 
The failure of gauge coupling 
unification in the standard model forces the SU(2)$_R$ breaking scale 
$M_R$ responsible for righthanded neutrino mass to lie much below 
$M_{GUT}$: $M_R\approx 10^{13}$ GeV. This in turn leads 
to too small righthanded neutrino masses: 
$max\left(m_{\nu_R}\right)\le 10^8$ GeV, since from $d=6$ 
proton decay constraints $M_{GUT}$ must definitely lie 
above $10^{15}$ GeV. 

This won't do: light neutrino masses will become generically too 
large. A possible way out is to give up the predictability 
and simply fine-tune the Dirac neutrino masses through a complicated 
enough Yukawa sector. This would be against the 
the original motivation of calculating and predicting fermion masses 
and mixings. Furthermore, so light righthanded neutrinos seem to be 
in contradiction with leptogenesis constraints \cite{Hamaguchi:2001gw}. 
Instead it is much 
more natural to look for a theory with $M_R\approx M_{GUT}$, since the 
scope of our program is the implementation of the Witten's mechanism 
in the minimal and predictive scenario. 

Unification constraints then apparently imply 
low energy supersymmetry, which would kill the radiative effect. The 
way out of this impasse is quite unique: for the sake of the one-step 
GUT symmetry breaking one should have light 
gauginos and higgsinos and at the same time 
the supersymmetry breaking scale close to $M_{GUT}$ in order to be 
in accord with neutrino masses. 

\begin{figure}[h]
\includegraphics[width=0.70\linewidth]{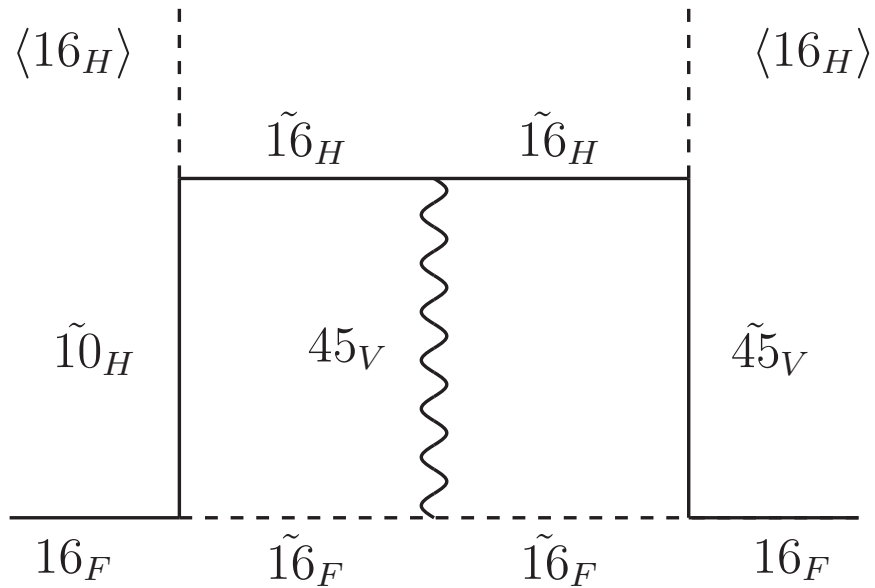}
\caption{\label{fig2} A supersymmetric 
contribution to the radiatively generated fermion mass. 
In our notation the tilde stands for the supersymmetric partners, 
i.e. $\tilde{45}_V$ denotes gauginos, $\tilde{16}_F$ squarks and 
sleptons and $\tilde{10}_H$ and $\tilde{16}_H$ higgsinos.}
\end{figure}

Thus we need to extend the original radiative 
mechanism to a (strongly broken) supersymmetric theory. In 
Fig. \ref{fig2} we give a typical contribution due to supersymmetric 
partners in the loops; the others are easily obtained. 

In the exact supersymmetric limit of course all the diagrams cancel 
against each other. Eq. (\ref{mnur}) gets simply traded for 

\beq
M_{\nu_R}\approx\left({\alpha\over\pi}\right)^2Y_{10}
{M_R^2\over M_{GUT}}f\left({\tilde{m}\over M_{GUT}}\right)\;,
\eeq

\noindent
where $\tilde{m}$ is the scale of supersymmetry breaking, or in other 
words the difference between the scalar and fermion masses of the 
same supermultiplet. This is valid only for $\tilde{m}$ not above 
$M_{GUT}$. The function $f(x)\to 0$ when $x\to 0$ and 
$f(x)={\cal O}(1)$ if $x={\cal O}(1)$.

Due to the two loops suppression the only way to have 
large enough righthanded neutrino masses is through single step 
symmetry breaking $M_R\approx M_{GUT}$ and the large 
$\tilde{m}\approx M_{GUT}$. Thus, independently of the details 
of the realistic Yukawa sector, one is forced to the split 
supersymmetry picture. 

If we keep only one $10_H$, we will of course have $m_D=m_L$ 
and $m_U=m_{\nu_D}$ for all three generations and the vanishing mixing 
angles. This is due to the well known quark-lepton symmetry of the 
$10_H$ vev being in the (2,2,1) of the PS symmetry. 
As a remedy we offer two simple possibilities. The first one 
uses another $10_H$ and the second one interchanges it for $120_H$. 
We describe them now in more detail.

\subsection{Model A}


Add another $10_H$; this allows for nonvanishing mixings since up 
and down fermion mass matrices are not anymore proportional to each other. 
At first glance, though, the above problem of equal down quark and charged 
lepton masses persists. There is a nice way out however: a radiatively 
induced (2,2,15) component of the effective $\overline{126}$ through the 
two loop diagrams as before, but with light fermions as external states 
and a small (order electroweak scale) vev of the $16_H$:

\beq
\label{mf}
M_f\approx\left({\alpha\over\pi}\right)^2\left(c_1Y_{10}^{(1)}
+c_2Y_{10}^{(2)}\right){M_RM_Z\over M_{GUT}}
g\left({\tilde{m}\over M_{GUT}}\right)\;,
\eeq

\noindent
where $c_i$ contain various numerical factors from the above diagrams 
and the mixings between between the SU(2)$_L$ doublets in $10_H$, 
and $16_H$, while $g(x)$ has similar properties 
as $f(x)$ for $x$ close to zero and of order $1$. These mixings arise 
from the interactions in the superpotential

\beq
W_H=\alpha_i 16_H 10_H^i 16_H \;.
\eeq

The contribution (\ref{mf}) by itself would imply $m_\mu=-3m_s$ at 
$M_{GUT}$, which works very well after being run down to $M_Z$. 
We thus propose this radiative mechanism as a possible natural 
way to obtain correct mass relations for charged fermions. 
There is more to it: unless there is low-energy 
supersymmetry such effects should be taken into account even in 
models that apparently work at the tree level (for example, see 
below the discussion of model B). 

Admittedly, a conspiracy between the tree level and the two loop 
contributions is needed in order to achieve correct relations for the 
first two generations. At the same time the gauge coupling at the GUT 
scale must be large enough: $(\alpha/\pi)^2> 10^{-3}$ or so, in order 
for the muon and the strange quark to weigh enough. This requires 
the existence of complete SU(5) multiplets at an intermediate scale 
and is naturally present in many models of the mediation of supersymmetry 
breaking. The appealing feature of this is an enhancement of the 
$d=6$ proton decay rate which can make proton decay observable in the 
near future; see the last reference in \cite{Arkani-Hamed:2004fb}. 
Recall that $d=5$ proton decay is negligible in this version of 
the split supersymmetry with sfermion masses at the GUT scale. 
In view of this a detailed analysis of different channel branching 
ratios of $d=6$ proton decay along the lines of 
\cite{FileviezPerez:2004hn} is called for.

On top of that, the 
righthanded neutrino mass matrix must be presumably rather hierarchical 
in order to compensate for a tree level hierarchy in $M_{\nu_D}$. 
Obviously a careful numerical anaysis is needed at this point, but 
the challenge is highly nontrivial and is beyond the scope of this 
letter. This is similar (and even more constrained) to the situation 
encountered in the type I seesaw case in the minimal SO(10) with 
$10_H$ and $\overline{126}_H$ case. There the type II seesaw 
\cite{Lazarides:1980nt} works very 
well and offers a natural connection between $b-\tau$ unification and 
the large atmospheric mixing angle \cite{Bajc:2002iw,Bajc:2001fe}. 
Here the type II 
contribution, although present, is strongly suppressed. It originates 
from the same type of diagrams as $M_{\nu_R}$, when the vevs of $16_H$ 
point in the SU(2)$_L$ rather than SU(2)$_R$ direction. While the two loop 
suppression of $M_{\nu_R}$ enhances the type I contribution to the 
seesaw formula, the same loop effect basically kills the type II effect. 

It is worth mentioning that $b-\tau$ unification is natural in this 
approach due to the tree level dominance of the $10_H$ Higgses. 
Furthermore, the model has the same small number of Yukawa couplings 
as the minimal renormalizable model with $10_H$ and 
$\overline{126}_H$: $15=3+6\times 2$ real parameters 
\cite{Aulakh:2003kg}.

\subsection{Model B}


Instead of another $10_H$ one can add a $120_H$ representation. 
Although a larger representation, it has even less Yukawa couplings, 
due to its antisymmetric nature in generation space: $9=3+3\times 2$ 
real parameters. The charged fermion masses with 
$10_H$ and $120_H$ have been studied both analytically and numerically 
in \cite{Matsuda:2000zp} for even more restrictive choice of parameters. 
The preliminary study indicates that the theory can work, but we 
believe that more detailed study is needed, especially since the 
neutrinos were not included. Some of the effects of $120_H$ were 
also studied in a model with $10_H$ and $\overline{126}_H$ Higgses 
as a subleading effect \cite{Bertolini:2004eq} and for 
the choice of type II seesaw. Here thus there is an interesting double 
challenge of less parameters and no choice for the type of seesaw: it 
must be type I as we stressed above. 

At first glance in this case loops seem irrelevant for the charged 
fermion massses, since there is $(2,2,15)$ effect already at the 
tree level. However its contribution is antisymmetric in generation 
space since it originates from $120_H$. Thus the same two loop effects 
as in the model A that generate a symmetric $(2,2,15)$ in the effective 
$\overline{126}_H$ must be included when a careful numerical analysis is 
performed. In this case there are additional diagrams where the 
external $16_H16_H$ (which generates an effective $\overline{126}_H$) 
are traded for say $120_H45_H$ or $120_H210_H$.

\subsection{Some phenomenological issues}

Obviously with scalar masses at $M_{GUT}$ the $d=5$ proton decay
operators become completely negligible (it is amusing that even
the possible $d=4$ operators in this case become harmless).    
In model A the usual $d=6$ gauge boson induced proton decay
is necessarily enhanced by a larger gauge coupling and thus likely
observable in the next generation of proton decay experiments 
\cite{jung}.  
In model B this depends on whether or not there are extra complete  
multiplets at some intermediate scale. Model A is further characterized by
symmetric Yukawa couplings. In this way one can obtain interesting
relations among different decay channels \cite{FileviezPerez:2004hn}.

The main characteristic of the split supersymmetry is the cosmologically 
stable lightest neutralino as the dark matter candidate and a long lived 
gluino. Gluino lifetime is given by 

\beq
\tau({\rm gluino})= 3.10^{-2}{\rm s}\left({\tilde{m}\over 10^9
{\rm GeV}}\right)^4\left({1 {\rm TeV}\over m_{gluino}}\right)^5\;.
\eeq

With $\tilde{m}$ bigger than $10^{15}$ GeV as in this theory gluinos 
lighter than $10$ TeV would be cosmologically stable. If gluinos 
form heavy nuclei, which seems plausible (for a recent analysis see 
\cite{Antoniadis:2004dt} and references therein), such nuclei should 
have been discovered by now. The lack of such evidence is normally 
attributed to gluino decay. In our case, this would imply gluino mass 
above $10$ TeV, completely out of LHC reach. Furthermore, one must 
make sure that gauge couplings still unify, not impossible due to 
possible GUT scale threshold effects. Another possibility is 
to appeal to a low reheat temperature after inflation so that 
gluinos are not produced (or they are washed out). 

One of the appealing aspects of low energy or split supersymmetry is 
the possibility of a neutralino being a dark matter candidate. If 
gluinos are really stable and thus need to be washed out, 
one must make sure that the LSP neutralino is not washed out at 
the same time. This would put an interesting constraint on inflation 
model building. 

\section{Conclusions and outlook}

In this letter we made a strong case for the radiative seesaw mechanism. 
The simplicity and the elegance of this approach makes it definitely worth 
reviving. We find that the price that needs to be paid to make it work 
is actually very low: it may be possible to add just another Higgs 
multiplet, either $10_H$ (model A) or $120_H$ (model B). Admittedly more 
work is needed to be sure that either of these models actually fits 
all the low-energy data; otherwise it may be necessary to complicate 
further the Yukawa sector. 

We also argued that similar radiative effects play an important role 
for light charged fermion masses. Such effects are necessarily present 
in the theories with radiative seesaw and they may even provide a cure 
for the wrong GUT scale relations in the minimal theory. In particular, 
if the Yukawa sector contains only $10_H$'s, these effects may be 
sufficient for having correct strange quark and muon masses (and 
certainly for down quark and electron masses). 

This paves way for new class of highly predictive and simple SO(10) 
models. The immediate important consequence is that supersymmetry must 
be broken at the GUT scale, but with light gauginos and higgsinos. 
Our work provides simultaneously a strong rationale for 
both radiative seesaw mechanism and split supersymmetry. What makes it 
particularly appealing is that both scenarios are potentially testable 
in the near future. 

{\it\bf Acknowledgments}\hspace{0.5cm}
We thank C. Aulakh, G. Dvali, A. Melfo, R. Mohapatra and F. Vissani 
for valuable comments and discussions. 
The work of G.S. was supported by EEC (TMR contracts ERBFMRX-CT960090 
and HPRN-CT-2000-00152), the work of B.B. by the Ministry of Education, 
Science and Sport of the Republic of Slovenia. B.B. thanks 
ICTP for hospitality during the course of this work. 

{\it\bf Note added}\hspace{0.5cm} 
After this work was completed a new paper \cite{Desai:2004xf} 
appeared which discusses radiative generation of fermion masses, 
but in a quite different appoach (utilizing singlet fermions). 
This paper also contains references to earlier works in the field.

\end{document}